# Inverse Confounding Analysis: An Exact Method for Quantifying the Significance of Confounding

**Sergey Porotsky, PhD**

**E-mail: sp0912@hotmail.com**

***Abstract.*** The presence of unmeasured confounding factors during the collection of observational data may lead to biased estimates of the effect of an exposure on an outcome. Consequently, a central problem in causal inference based on observational data is sensitivity analysis with respect to unmeasured confounding. Existing sensitivity analyses generally focus on worst-case bounds. We propose an exact method for quantifying the significance of confounding, defined here in terms of the complete range of analytical estimates of the stratification-based Risk Ratio over the set of all joint distributions compatible with the observed characteristics. We refer to the proposed method as **Inverse Confounding Analysis (ICA).**

The proposed ICA method extends the widely used E-value approach but, in contrast to it, does not restrict the analysis to a worst-case lower bound. Instead, it provides exact estimates over the entire set of admissible configurations. This requires several additional input parameters, namely the frequencies of the exposure, the confounder, and the outcome. The ICA method is based on an inverse problem: reconstructing the set of admissible joint distributions from specified frequencies and pairwise associations. We formulate this reconstruction problem as a system of nonlinear equations and obtain an analytical solution. Surprisingly, the complete solution set can be parameterized linearly by a single free parameter. The corresponding stratification-based Risk Ratio is then represented as a fractional-linear function of this parameter. This representation makes it possible to derive exact analytical measures of the significance of confounding over the entire set of admissible statistical configurations.

## 1. Introduction

There are two fundamentally different approaches to collecting statistical data:

- randomized controlled trials;
- observational (passive) data collection.

The first approach underlies, for example, the methodology used by the FDA for the evaluation of new drugs in Phase III clinical trials. When properly designed and conducted, randomization provides substantially less confounded estimates of the effect of the investigated factor. Random assignment of the original cohort to the treatment and control groups breaks systematic associations between treatment assignment and other factors, including both measured and unmeasured baseline factors. Unfortunately, for economic and ethical reasons, randomized controlled trials cannot be used for many practical research questions. Most statistical studies therefore rely on observational data, and this paper focuses on this setting.

In recent years, numerous results based on observational data have been reported. For example, it may be found that the use of a particular food, beverage, dietary supplement, medication, or treatment method is associated with a 20%, 30%, or 40% reduction or increase in some outcome. The outcome may be the incidence of a disease, the probability of successful treatment, mortality, or another event. Such associations are often interpreted as evidence of a substantial effect of the investigated factor. However, it is entirely possible that another factor simultaneously affects both the investigated exposure and the outcome. In such a case, the observed association between the exposure and the outcome may not represent a causal effect.

When the exposure E and outcome D are binary variables, the Risk Ratio (RR) is commonly used to quantify their association. Suppose that, in a statistical cohort:

- the exposure E takes the value 1 in M1 observations, among which the outcome D takes the value 1 in N1 observations;
- the exposure E takes the value 0 in M0 observations, among which the outcome D takes the value 1 in N0 observations.

Then RR = (N1 /M1)/(N0 /M0).

In probability notation, RR = Prob(D = 1/E = 1)/ Prob(D = 1/E = 0).

A high Risk Ratio between E and D does not, by itself, establish that E causally affects D. A large RR may occur even when there is no causal relationship between the two variables. We refer to such an association as a spurious association. There are several methods for investigating whether an observed association may be explained by confounding. For discrete factors, stratification is one of the classical approaches [1]. The basic idea is to divide the original cohort into subcohorts in which a suspected confounder U has a fixed value, and then analyze the data separately within each stratum.

For example, a primary analysis of IVF data [2, 3] produces the seemingly surprising result that women aged 45–50 years have a higher IVF success rate than women aged 43–44 years. For the age factor, the observed Risk Ratio is RR = 1.42. This result appears counterintuitive. Stratification was therefore used to investigate a number of potential confounders recorded in the study, including the number of previous IVF procedures, type and cause of infertility, stimulation method, oocyte source, sperm source, and others.

The analysis showed that the apparent association was explained by the oocyte source — whether the patient's own oocyte or a donor oocyte was used. Women aged 45–50 years were substantially more likely to use donor oocytes, which are generally of higher quality, than women aged 43–44 years. After analyzing the data separately within the patient-oocyte and donor-oocyte strata, the apparent paradox disappeared. The resulting stratification-based Risk Ratio was 0.76, indicating that younger women have a higher IVF success rate.

Throughout this paper, the term **stratification-based Risk Ratio** refers specifically to the Risk Ratio obtained after stratification by the confounder under investigation. It does not denote an absolute or uniquely defined causal effect.

Stratification and similar methods, however, can address confounding only by factors that were recorded during data collection. They cannot directly address unmeasured factors.

For example, [4] investigated the association between coffee preparation method (filtered versus unfiltered) and mortality. The following factors were measured: age, number of cigarettes smoked per day, cholesterol, triglycerides, blood pressure, body mass index, education, and year of examination. The researchers concluded: “Well-known factors that could confound the results of the study were measured and analyzed. We found higher mortality from cardiovascular disease in the unfiltered coffee drinker group...” [4]. However, potentially important unmeasured characteristics may still exist. For example, suppose that a person's temperament affects both the choice of coffee preparation method and health-related behavior. A strong association between coffee preparation and mortality could then arise even without a causal effect of coffee preparation on mortality. Similarly, [5] investigated the association between sleep-onset timing and cardiovascular disease and stratified the analysis by factors such as age, sex, obesity, cholesterol, hypertension, diabetes, and hypotension. An unmeasured factor such as temperament could nevertheless potentially influence both sleep habits and cardiovascular risk. The same general issue occurs in other observational studies. For example, [6] reported an association between previous influenza vaccination and the probability of developing COVID-19. A potentially important factor such as personal health-related discipline might influence both the decision to receive influenza vaccination and behaviors such as mask use and social distancing. If so, part of the observed association could be due to confounding.

These examples illustrate a general methodological problem rather than necessarily indicating flaws in the individual studies. In principle, reliable causal conclusions from observational data require adequate control of factors that jointly influence the exposure and outcome.

The most reliable way to achieve this is randomization. Randomization statistically balances both observed and unobserved factors between treatment groups. In some relatively rare situations, randomization can also arise naturally in observational settings. One example is described in [7], where military service during the Vietnam War was assigned using a draft lottery. The resulting external randomization enabled researchers to estimate the effect of military service on various socioeconomic outcomes. Instrumental-variable methods may provide another way of addressing unmeasured confounding [8]. The basic idea is to replace the analysis of the exposure itself — which may be associated with the outcome because of hidden factors — with an analysis based on another variable, the instrument. The approach attempts to reproduce a sampling mechanism analogous to a randomized experiment through an external variable, provided that a suitable instrument can be found. The key assumption is that the instrumental variable has no direct effect on the outcome and acts exclusively through the exposure under investigation. This requirement is also the most vulnerable element of the instrumental-variable approach.

Several other sensitivity-analysis methods have also been developed. One of the best-known is the E-value method, which is discussed in the next section.

The remainder of this paper is organized as follows:

- Section 2 reviews the E-value method and discusses several features that may lead to ambiguous interpretations of its results;
- Section 3 formulates the inverse reconstruction problem based on pairwise associations and frequencies of the exposure E, confounder U, and outcome D; analytical expressions are derived for the complete set of admissible joint distributions and the corresponding stratification-based Risk Ratios;
- Section 4 develops distribution-independent and distribution-dependent measures of Confounding Significance;
- Section 5 illustrates the proposed ICA method using published observational data and compares its results with those obtained using the E-value approach;
- Section 6 summarizes the principal results and possible extensions.

## 2. Analysis of the E-value Method

The E-value method was developed to address, at least in part, the problem of assessing the potential influence of an unmeasured confounder on an observed association [9–11]. The method has subsequently been widely applied and cited [12–14]. The following discussion considers the binary case and assumes, for simplicity, that the relevant Risk Ratios are greater than or equal to one. Analogous arguments can be formulated for Risk Ratios below one by appropriately reversing the coding of the binary variables. Let

- RR_ED denote the association between E and D;
- RR_EU denote the relevant measure of association between E and U;
- RR_UD denote the relevant measure of association between U and D.

The E-value literature emphasizes « In practice, we do not know the strengths of the unmeasured confounder associations, but we could, in principle, specify many different values and determine how the estimate is affected by each setting» [9].

The key sensitivity-analysis result underlying the E-value method [9, 10] is that, under the corresponding assumptions, the confounding-induced distortion of the observed Risk Ratio is bounded by

- Bias = (RR_EU*RR_UD)/( RR_EU + RR_UD – 1). (1)

Let RR_ED_obs denote the Risk Ratio obtained by directly analyzing the observed data without adjustment for U.

In this paper, let RR_ED_strat denote the Risk Ratio obtained after stratification by U.
Thus, RR_ED_strat >= RR_ED_obs/Bias [9].
In [9], the E-value method is illustrated using the association between infant feeding type (breastfeeding versus formula feeding) and mortality from respiratory disease. The observed Risk Ratio was RR_ED_obs = 3.9. Smoking was considered as a potential confounder. Suppose that the association between smoking and respiratory mortality is assumed to be at most RR_UD = 4, and the association between smoking and breastfeeding status is assumed to be at most RR_EU = 2. Because the lower bound exceeds one, the assumed strength of smoking-confounding is insufficient to completely eliminate the observed association.
When plausible maximum values of RR_EU and RR_UD cannot be specified, [9] proposes considering the set of pairs satisfying
(RR_EU*RR_UD)/( RR_EU + RR_UD – 1) = RR_ED_obs.
In the special case, when RR_EU = RR_UD, this gives value
RR_ED_obs + sqrt( RR_ED_obs*(RR_ED_obs – 1) ), which is named the E-value.
Two features of the E-value framework are particularly relevant to the proposed Inverse Confounding Analysis (ICA) method.

## 2.1 One-sided nature of the E-value bound

The E-value bound is inherently one-sided. If Bias < RR_ED_obs, then RR_ED_strat > 1, and the considered confounder cannot completely eliminate the observed positive association.
If, however, Bias > RR_ED_obs , the bound no longer establishes whether the stratification-based Risk Ratio is below or above one – conclusion that RR_ED_strat < 1 will be unclear. The appropriate conclusion is therefore only that the confounder **could** potentially explain the observed association. This feature is illustrated on chapters 5.1 and 5.3.
Importantly, the E-value calculation does not quantify how large a proportion of the admissible underlying statistical configurations would actually lead to RR_ED_strat < 1. This distinction is central to the present ICA method.

## 2.2 The E-value definitions of RR_EU and RR_UD

A potentially important distinction is that the quantities RR_EU and RR_UD used in the E-value formula are not necessarily identical to the Risk Ratios that would be calculated from the observed data using the standard Risk Ratio formula.
For RR_EU, [10] defines

- RR_EU = max{ RR_EU_U1, RR_EU_U0}, where
- RR_EU_U1 = Prob( U= 1/E = 1)/ Prob(U = 1/E = 0);
- RR_EU_U0 = Prob( U= 0/E = 1)/ Prob(U = 0/E = 0);

The observed Risk Ratio is defined as RR_EU_obs = Prob( U= 1/E = 1)/ Prob(U = 1/E = 0). If, according our assumption, RR_EU_obs > 1, then RR_EU_U0 < 1, and therefore RR_EU = RR_EU_obs.

Thus, in the present setting, replacing the E-value method RR_EU by the corresponding observed Risk Ratio does not introduce an additional discrepancy.

The situation is different for RR_UD. The E-value definition considers the maximum relative difference between the risks of D across the levels of U within each level of E:

- RR_UD = max{ RR_UD_E1, RR_UD_E0}, where
- RR_UD_E1 = max{ Prob( D = 1/E = 1, U = 1), Prob(D = 1/E = 1, U = 0) }/ min{ Prob( D = 1/E = 1, U = 1), Prob(D = 1/E = 1, U = 0) }
- RR_UD_E0 = max{ Prob( D = 1/E = 0, U = 1), Prob(D = 1/E = 0, U = 0) }/ min{ Prob( D = 1/E = 0, U = 1), Prob(D = 1/E = 0, U = 0) }

In contrast, the Risk Ratio from observations is

- RR_UD_obs = Prob( D = 1/U = 1)/ Prob(D = 1/U = 0).

In general, it is entirely possible that RR_UD > RR_UD_obs - this situation occurs in all three practical examples considered in Section 5. Consequently, substituting the observed RR_UD_obs for the method E-value calculated RR_UD may underestimate the Bias factor and may therefore lead to a different interpretation of the potential confounding. So, if observational data are used for Bias calculation, even in the case of Bias < RR_ED_obs, a conclusion that RR_ED_strat > 1 will be unclear.

This distinction is not merely terminological: it can materially affect the assessment of whether a confounder is capable of explaining an observed association. It is iluustrated on chapter 5.2

The proposed ICA method addresses a broader question than the E-value method. Rather than estimating only a worst-case lower bound, it reconstructs the complete set of admissible joint distributions compatible with the observed characteristics and calculates the corresponding stratification-based Risk Ratios.

## 3. Proposed Inverse Confounding Analysis (ICA) Method

To address the limitations discussed above, we propose a method that extends the E-value framework but focuses on the complete set of admissible configurations rather than on a single worst-case configuration. Instead of asking only “Can there exist a confounder such that the distortion of the observed Risk Ratio reverses the result?” the proposed ICA method addresses the broader question: “What proportion of the admissible space of statistical configurations leads to the disappearance of the observed effect?”

As noted in Section 2, the E-value method can establish, under the relevant assumptions, that a specified confounder cannot completely eliminate the observed association. The opposite situation — whether the confounder actually does eliminate the association and how often this occurs over the admissible set — is not quantified by the E-value bound.

We therefore formulate the problem as an inverse problem: given frequencies and pairwise association measures, reconstruct the set of joint distributions compatible with these quantities.

The general idea of reconstructing an unknown joint distribution from a set of known distributional characteristics has parallels with the classical method of moments. However, the present approach is conceptually different: the constraints used here are not moments, and no parametric family of distributions is assumed. Instead, the joint probabilities Pijk are reconstructed directly from empirically specified probabilities and pairwise Risk Ratios. We refer to this approach as Inverse Confounding Analysis (ICA).

The proposed ICA method requires some additional information compared with the E-value method, namely the frequencies of the exposure, confounder, and outcome.

Let :

- Pijk = Prob(E = i, U = j, D = k), where

i = 0, 1; j = 0, 1; k = 0, 1.

Remember that RR_ED_obs denote the observed Risk Ratio obtained without adjustment for U, and RR_ED_strat denote the Risk Ratio obtained after stratification by U.

For the stratification-based Risk Ratio we use expression

- RR_ED_strat = (p_str1*Up1+p_str0*Up0)/(p_str1*Dn1+p_str0*Dn0), (2)

where :

p_str1 = p111 + p110 + p011 + p010 , is the probability of stratum (U = 1),

p_str0 = p101 + p100 +p001 + p000, is the probability of stratum (U = 0),

Up1 = p111/(p110 + p111); Up0 = p101/(p100 + p101);

Dn1 = p011/(p010 + p011); Dn0 = p001/(p000 + p001);

If the complete three-way table of E, U, and D were available, all eight probabilities could be calculated directly and equation (2) could be evaluated. This is the usual situation in a classical stratification problem. For example, in the IVF study discussed in the Introduction, the complete statistical information for 8886 observations is available [2]. In the unmeasured-confounding problem considered here, however, we assume that the complete three-way table is unavailable and that only pairwise associations are known.

We therefore formulate the inverse problem of reconstructing the joint distribution of the eight probabilities.

Suppose that we additionally know

pE = Prob(E = 1)

pU = Prob(U = 1)

pD = Prob(D = 1).

The values of pE and pD can normally be obtained directly from the observed data. The value of pU may be more difficult to determine when U is unmeasured, but it may nevertheless be substantially easier to estimate than the pairwise Risk Ratios involving U.

The reconstruction problem is therefore:

Given pE, pU, pD, RR_ED_obs, RR_EU_obs and RR_UD_obs find the complete solution set of the following seven nonlinear equations in eight unknowns:

p111 +p110 + p101 + p100 + p011 + p010 + p001 + p000 = 1; (3)

p100 + p101 + p110 + p111 = pE; (4)

p010 + p110 + p011 + p111 = pU; (5)

p001 + p011 + p101 + p111 = pD; (6)

( (p101 + p111)/(p100 + p101 + p110 + p111) )/( (p001 + p011)/(p000 + p001 + p010 + p011) ) = RR_ED_obs; (7)

- from expression RR_ED_obs = Prob(D = 1/E = 1)/ Prob(D = 1/E = 0).

( (p110 + p111)/(p100 + p101+ p110 + p111) )/( (p010 + p011)/(p000 + p001 + p010 + p011) ) = RR_EU_obs; (8)

- from expression RR_EU_obs = Prob(U = 1/E = 1)/ Prob(U = 1/E = 0).

( (p011 + p111)/(p010 + p110 + p011 + p111) )/( (p001 + p101)/(p000 + p001 + p100 + p101) ) = RR_UD_obs; (9)

- from expression RR_UD_obs = Prob(D = 1/U = 1)/ Prob(D = 1/U = 0).

subject to :

0 <= Pijk <= 1 (i, j, k = 0, 1) (10)

Because there are eight unknown probabilities and seven independent constraints, the solution set is generically expected to be one-dimensional.

**Proposition 1. Invariance of the solution set**

If {p111, p110, p101, p100, p011, p010, p001, p000} is a solution of equations (3)–(9), then, for any value of q the probabilities {p111 + q, p110 - q, p101 - q, p100 + q, p011 - q, p010 + q, p001 + q, p000 - q} is also a solution.

The proposition follows directly by substitution into equations (3) – (9).

Define:

C110 = (pE*pU)/((1 - pE)/RR_EU_obs + pE),

C101 = (pE*pD)/((1 - pE)/RR_ED_obs + pE),

C100 = pE - (pE*pD)/((1 - pE)/RR_ED_obs + pE) - (pE*pU)/((1 - pE)/RR_EU_obs + pE),

C011 = (pU*pD)/((1 - pU)/RR_UD_obs + pU),

C010 = pU - (pE*pU)/((1 - pE)/RR_EU_obs + pE) - (pU*pD)/((1 - pU)/RR_UD_obs + pU),

C001 = pD - (pU*pD)/((1 - pU)/RR_UD_obs + pU) - (pE*pD)/((1 - pE)/RR_ED_obs + pE),

C000 = 1 + (pE*pD)/((1 - pE)/RR_ED_obs + pE) + (pE*pU)/((1 - pE)/RR_EU_obs + pE) + (pU*pD)/((1 - pU)/RR_UD_obs + pU) - (pE + pU+ pD).

### Proposition 2. Definition of the complete solution set

The complete solution set of equations (3) – (9) can be defined by p111 as follows:

- p110 = C110 - p111; (11)
- p101 = C101 - p111; (12)
- p100 = C100 + p111; (13)
- p011 = C011 - p111; (14)
- p010 = C010 + p111; (15)
- p001 = C001 + p111; (16)
- p000 = C000 - p111; (17)

Although the inverse problem is nonlinear in its original formulation, its complete solution set therefore admits a simple linear representation.

**Proof.** Equations (3)–(6) can be rewritten as

p100 = pE - p101 – p110 - p111;

p010 = pU - p110 - p011 - p111;

p001 = pD - p011 - p101 - p111;

p000 = 1- (pE + pU +pD - p110 - p101 - p011 – 2*p111);

Expanding equation (7) gives

RR_ED_obs = (p101 + p111)/pE = RR_ED_obs *(pD – p101 – p111)/(1 – pE), and therefore

p101 = (pE*pD)/((1 - pE)/RR_ED _obs + pE) – p111.

Analogously, equations (8) and (9) give the corresponding expressions for p110 and p011, from which equations (13), (15), (16), and (17) follow.

It follows that:

- the solution set of the original nonlinear system is a straight line in the eight-dimensional space of the variables Pijk;

- as p111 increases, p110, p101, p011 and p000 decrease monotonically, whereas p100, p010, and p001 increase monotonically;
- after imposing the probability constraints (10), the complete admissible solution set is an interval p111 = [p111_low…p111_high]

For the decreasing variables p110, p101, p011, p000:

- the constraints Pijk >= 0 give Cijk – p111 >= 0, and so p111 <= min(1, C110, C101, C011, C000).
- the constraints Pijk <= 1 give Cijk – p111 <= 1, and so p111 >= max(0, C110 - 1, C101 - 1, C011- 1, C000 - 1).

For the increasing variables p100, p010, p001:

- the constraints Pijk >= 0 give Cijk + p111 >= 0, and so p111 >= max(0, -C100, -C010, -C001).
- the constraints Pijk <= 1 give Cijk + p111 <= 1, and so p111 <= min(1, 1 - C100, 1 - C010, 1 - C001).

Consequently :

- p111_low = max(0, C110 - 1, C101 - 1, C011- 1, C000 – 1, -C100, -C010, -C001) (18)
- p111_high = min(1, C110, C101, C011, C000, 1 - C100, 1 - C010, 1 - C001) (19)

**Compatibility criterion.** If p111_low > p111_high, the system of equations (3) – (9) has no solution satisfying the probability constraints (10). Thus, there is no joint distribution of E, U, and D that is compatible with the specified values of pE, pU, pD, RR_ED_obs, RR_EU_obs, RR_UD_obs. The proposed method therefore also provides a feasibility test for the supplied input characteristics. We refer to every solution satisfying equations (3)–(10) as an admissible configuration of the original probabilities.

## Proposition 3. Main result

The stratification-based Risk Ratio can be represented as

**RR_ED_strat = (a +в*p111)/(c + d*p111),** (20)

where a, b, c, and d depend only on the input parameters pE, pU, pD, RR_ED_obs, RR_EU_obs, RR_UD_obs. Specifically,

a = (C000 + C001)*(C010 + C011)*C110*C101*(C000 + C001 + C100 + C101);

b = (C000 + C001)*(C010 + C011)*((C100 + C101)*(C010 + C011 + C110) - C110*(C000 + C001 + C100 + C101));

c = C110*(C100 + C101)*(C001*(C010 + C011)*(C000 + C001 + C100 + C101)+ C011*(C000 + C001)*(C010 + C011 + C110));

d = C110*(C100 + C101)*((C010 + C011)*(C000 + C001 + C100 + C101) - (C000 + C001)*(C010 + C011 + C110));

The proposition follows by substituting equations (11) – (17) into equation (2) and collecting like terms.

## Proposition 4. Monotonicity

The function RR_ED_strat(p111) is monotonic over the admissible interval.

**Proof.** The derivative of equation (20) is (b*c – a*d)/( (c + d*p111)^2 ).

On the admissible interval, the denominator is positive wherever the Risk Ratio is defined, while the numerator does not depend on p111. Hence the derivative has a constant sign throughout the admissible interval.

# 4. Consequences of the Main Result

The Main Result represents the stratification-based Risk Ratio RR_ED_strat as a fractional-linear function of a single free parameter p111. The consequences can therefore be divided into two groups:

- results independent of the assumed distribution of p111;
- results depending on the assumed distribution of p111.

## 4.1 Results Independent of the Assumed Distribution of p111

Consider equation RR_ED_strat(p111) = 1 (21)

If b =/= d, the unique solution is

- p111_root = (c - a)/(b - d) (22)

If RR_ED_strat(p111_low) < RR_ED_strat(p111_high), the function RR_ED_strat(p111) increases monotonically over the admissible interval. Values below p111_root correspond to RR_ED_strat < 1, whereas values above p111_root correspond to RR_ED_strat > 1.

If RR_ED_strat(p111_low) > RR_ED_strat(p111_high), the function RR_ED_strat(p111) decreases monotonically, and the corresponding inequalities are reversed : values below p111_root correspond to RR_ED_strat > 1, whereas values above p111_root correspond to RR_ED_strat > 1.

Thus, instead of asking only “Can there exist a confounder for which the stratification-based Risk Ratio is below one?” proposed ICA method can answer the broader question:

“What proportion of the admissible space of statistical configurations leads to a stratification-based Risk Ratio below one?”

The proposed ICA method also provides both a lower and an upper bound.

Denote

- Max_RR_ED_strat  = max{ RR_ED_strat(p111_low), RR_ED_strat(p111_high) }  (23)
- Min_RR_ED_strat  = min{ RR_ED_strat(p111_low), RR_ED_strat(p111_high) },

Because the function is monotonic, Min_RR_ED_strat <= RR_ED_strat <= Max_RR_ED_strat. Thus, even if Max_RR_ED_strat > 1 and complete elimination of the observed positive association cannot be guaranteed for every admissible configuration, the stratification-adjusted effect is nevertheless bounded above by Max_RR_ED_strat.

In particular, if Max_RR_ED_strat < RR_ED_obs, the confounder cannot produce a stratification-based Risk Ratio as large as the observed unadjusted value. Example of such situation is considered on chapter 5.3.

The following distribution-independent quantities can therefore be obtained:

- the admissible interval [p111_low…p111_high];
- the location of p111_root, where RR_ED_strat = 1;
- the relative length of the region where RR_ED_strat < 1;
- the minimum and maximum possible values of RR_ED_strat.

## 4.2 Results Depending on the Assumed Distribution of p111

Once the admissible set of configurations and the analytical representation of RR_ED_strat are known, various measures can be constructed under an assumed distribution of p111, including

- Prob(RR_ED_strat < 1),
- Mean(RR_ED_strat),
- Median(RR_ED_strat), and etc.

The minimum and maximum possible values of RR_ED_strat can be determined from the available marginal information and the specified association measures. However, this information does not determine the probability distribution of p111 within the admissible interval. In particular, different distributions within the same interval may be compatible with exactly the same available information. Thus, the distribution of p111 is not identifiable from the available constraints alone. The available information determines the admissible interval [p111_low…p111_high], but does not provide any information that would justify assigning different probabilities to different values of p111 within this interval. There is no basis for assuming that one admissible value p111, or one subinterval of the admissible range, is more or less probable than another. So, in the absence of any additional information that would favor particular configurations, all admissible values of p111 are assigned equal value. Accordingly, we assume that p111 follows a uniform distribution over the admissible interval [p111_low…p111_high]. This assumption can be viewed as a maximum-entropy or non-informative principle: among all probability distributions supported on a finite interval, the uniform distribution introduces no preferential weighting within that interval .

Importantly, this assumption cannot be directly tested using the same information from which the admissible interval is derived. Such a test would require additional empirical information about the distribution of p111 within the admissible interval.

On the other hand, this assumption cannot be verified through an analysis of actual statistical data (i.e., complete tables containing values for factors E, U, and D), as the value p111 in such data is fixed and does not follow any specific distribution. Furthermore, even a Monte Carlo simulation would not allow for the assessment of this assumption, given that we have absolutely no knowledge of the distribution law according to which the input data should be generated.

As the main distribution-dependent measure of confounding significance, we propose the **Significance of Confounder**, denoted by **Sign_Conf**: the percentage of admissible configurations for which the stratification-based Risk Ratio satisfies RR_ED_strat < 1.

Under a uniform distribution of p111, monotonicity gives the following expressions.

If RR_strat(p111_low) > 1 & RR_strat(p111_high) > 1, Sign_Conf = 0 ; (24)

If RR_strat(p111_low) < 1 & RR_strat(p111_high) < 1, Sign_Conf = 100; (25)

If RR_strat(p111_low) >= 1 & RR_strat(p111_high) < 1, Sign_Conf = 100*(p111_high - p111_root)/(p111_high - p111_low) ; (26)

If RR_strat(p111_low) < 1 & RR_strat(p111_high) >= 1, Sign_Conf = 100*(p111_root - p111_low)/(p111_high - p111_low) ; (27)

The proposed ICA method can address not only the question “What proportion of admissible configurations leads to complete elimination of the observed E → D positive effect?”, but also the broader question “What proportion of configurations corresponds to a reduction of the observed effect by a factor of K? ”.

Suppose that we are interested in RR_ED_strat <= RR_ED_obs/K, where 1 < K <= R_obs.

Instead of equation (21), we solve RR_strat(p111) = R_obs/K,

The solution is p111_root = ( (R_obs/K)*c – a )/( b – (R_obs/K)*d ).

The subsequent formulas for calculating the corresponding proportion are analogous to equations (24) – (27).

## 5. Numerical Examples

We first illustrate the proposed Inverse Confounding Analysis (ICA) method and compare it with the E-value method using data from [15]. This study investigated, in a meta-analysis, the association between antidepressant use during pregnancy and the risk of miscarriage. The study was selected as the primary example because it is a well-known practical illustration of the E-value method, contains the required association measures, and therefore allows a direct comparison between the two approaches. The authors note that «Because both untreated depression and drug treatments might adversely affect pregnancy, treatment of depression in pregnant women involves balancing the benefits of treated depression against the potential treatment related risks to the mother and unborn child» [15].

Two potential confounders were considered: **smoking** and **alcohol consumption** during pregnancy. We use the following notation:

- E — antidepressant use;
- U — smoking or alcohol consumption;
- D — miscarriage.

The meta-analysis reported [15]:

- RR_ED_obs = 1.41,
- RR_EU_obs = 10.25 for alcohol consumption and RR_EU_obs = 2.06 for smoking,
- RR_UD_obs = 3.1 for alcohol consumption and RR_UD_obs = 1.32 for smoking.

## 5.1 Alcohol Consumption

For alcohol consumption, the E-value Bias calculation gives Bias = 2.6, which is substantially greater than RR_ED_obs = 1.41. The authors therefore conclude that «substantial alcohol use **could explain** the observed association between the use of antidepressants and risk of miscarriage». [15].

This conclusion is fully consistent with the one-sided nature of the E-value bound: RR_ED_strat >= RR_ED_obs/Bias.

However, the phrase "could explain" does not quantify how plausible such configurations among all statistical configurations compatible with the observed characteristics.

We therefore apply the proposed ICA method. As in the other examples below, we initially assume a Uniform distribution of p111 over its admissible interval. For illustration, we use

- pE = 0.1, pU = 0.15, pD = 0.1

Equations (18) – (27) give:

p111_low = 0.0, p111_high = 0.0135, p111_root = 0.0113, Min_RR_ED_strat = 0.23, Max_RR_ED_strat = 4.47.

The resulting dependence of the stratification-based Risk Ratio on p111 is shown in Fig. 1.

The proposed ICA method gives Sign_Conf = 16%. Thus, under the assumed Uniform parameterization, only 16% of the admissible configurations correspond to complete elimination of the positive association (RR_ED_strat < 1).

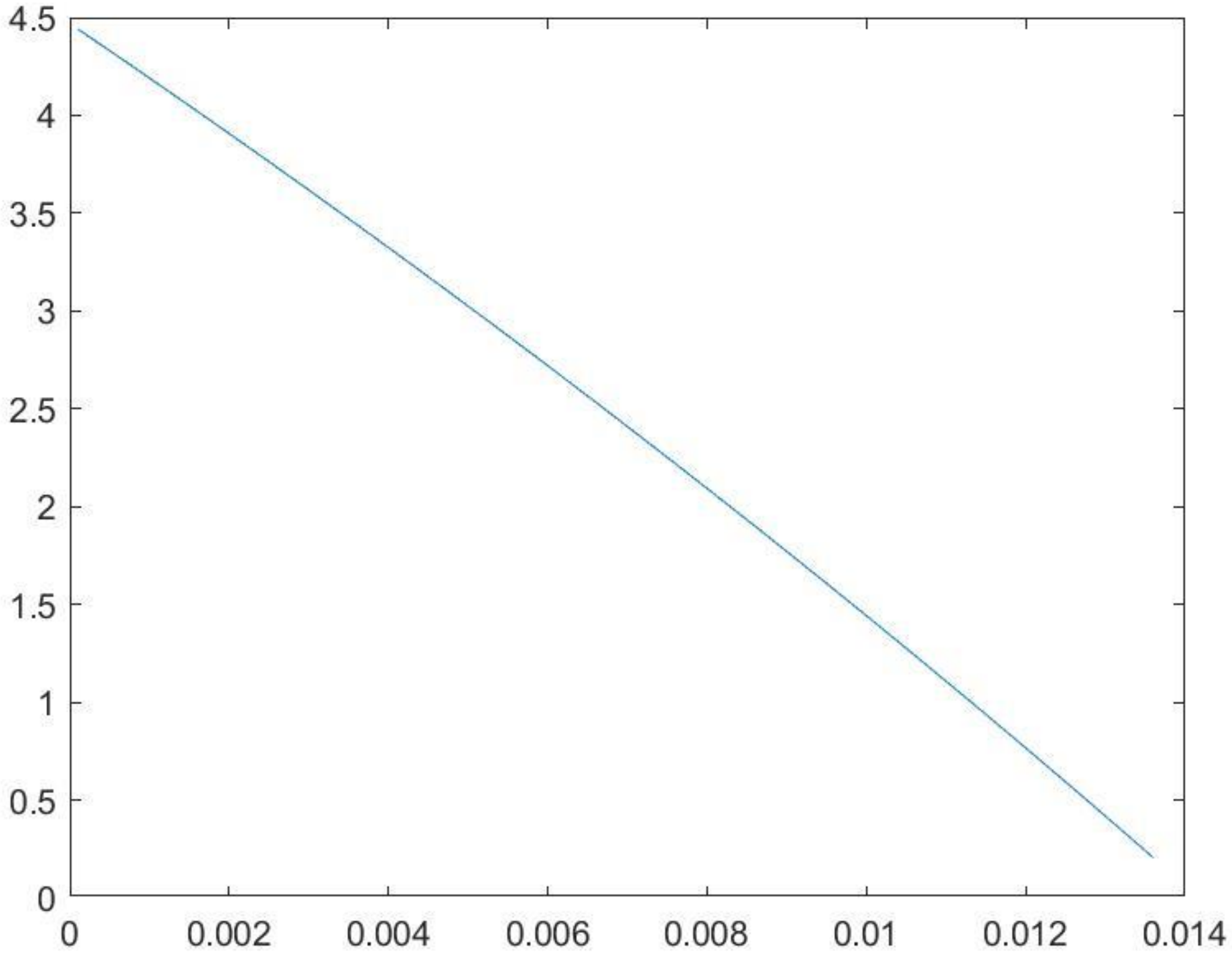

Figure 1. Dependence of the stratification-based Risk Ratio on p111 for the alcohol-consumption confounder

**Robustness to input uncertainty.** To assess robustness, we considered small perturbations of the input parameters corresponding to plausible measurement errors. The parameters pE, pU, pD, RR_ED_obs, RR_EU_obs and RR_UD_obs were independently perturbed by random errors uniformly distributed between −2% and +2% of their original values.

A Monte Carlo simulation with 100,000 runs gave MAE(Sign_Conf) = 0.8%, where MAE is «Mean Absolute Error». The resulting estimate of the Significance of Confounder therefore appears reasonably robust to the assumed measurement error.

### Comparison with a specific admissible configuration.

To illustrate the difference between the E-value bound and the proposed Inverse Confounding Analysis, consider a hypothetical dataset of 10,000 observations: N111 = 100, N110 = 699, N101 = 35, N100 = 165, N011 = 254, N010 = 448, N001 = 611, N000 = 7688, where Nijk denotes the number of observations with E = i, U = j, D = k (i, j, k = 0,1), Pijk = Nijk/10000.

Direct calculation of the stratification-based Risk Ratio using equation (2) gives RR_ED_strat = 1.43 > 1. Thus, for this particular admissible configuration, alcohol consumption does **not** completely eliminate the positive association between antidepressant use and miscarriage. This is entirely consistent with the E-value bound. The E-value result states only that the distortion cannot exceed the Bias factor; the actual distortion may be much smaller, and in this configuration it is essentially absent.

Among the admissible configurations, situations with RR_ED_strat > 1 account for 84%.

This result indicates that the statement that substantial alcohol use could explain the observed association should not be interpreted as evidence that such an explanation is typical among the admissible configurations

## 5.2 Smoking

For smoking, the E-value calculation gives Bias = 1.14, which is below RR_ED_obs = 1.41.

The E-value bound therefore gives RR_ED_strat >= RR_ED_obs/Bias > 1. On this basis, [15] concludes that “smoking is unlikely to explain the observed Risk Ratio of 1.41 on its own”.

We now apply the proposed ICA method. Using the same illustrative frequencies, pE = 0.1, pU = 0.15, pD = 0.1, equations (18) – (27) give p111_low = 0, p111_high = 0.0135, p111_root = 0.0100, Min_RR_ED_strat = 0.76, Max_RR_ED_strat = 1.65.

The dependence on p111 is shown in Fig. 2.

The proposed method gives Sign_Conf = 26%. Thus, 26% of the admissible configurations correspond to RR_ED_strat < 1.

A Monte Carlo simulation with 100,000 runs, using independent ±2% uniformly distributed perturbations of the input parameters, gave MAE(Sign_Conf) = 1.2%. The estimate is therefore reasonably robust to the assumed measurement error.

**A specific admissible configuration.** Consider a hypothetical dataset of 10,000 observations: N111 = 115, N110 = 164, N101 = 20, N100 = 700, N011 = 74, N010 = 1147, N001 = 791, N000 = 6989.

These data give RR_ED_obs = 1.41, RR_EU_obs = 2.06, RR_UD_obs = 1.32, and Bias = 1.14. So, RR_ED_strat >= RR_ED_obs/Bias > 1. However, direct calculation after stratification gives RR_ED_strat = 0.89. Thus, for this particular configuration, smoking completely eliminates the observed positive association.

This apparent discrepancy is explained by the distinction between the E-value definition of RR_UD and the observed Risk Ratio (RR_UD_obs). For this dataset, RR_UD_obs = 1.32, whereas the E-value definition gives RR_UD = 14.8. Using RR_UD = 14.8 instead of the value RR_UD_obs = 1.32 gives Bias = 1.92. The corresponding E-value bound is then compatible with RR_ED_strat < 1.

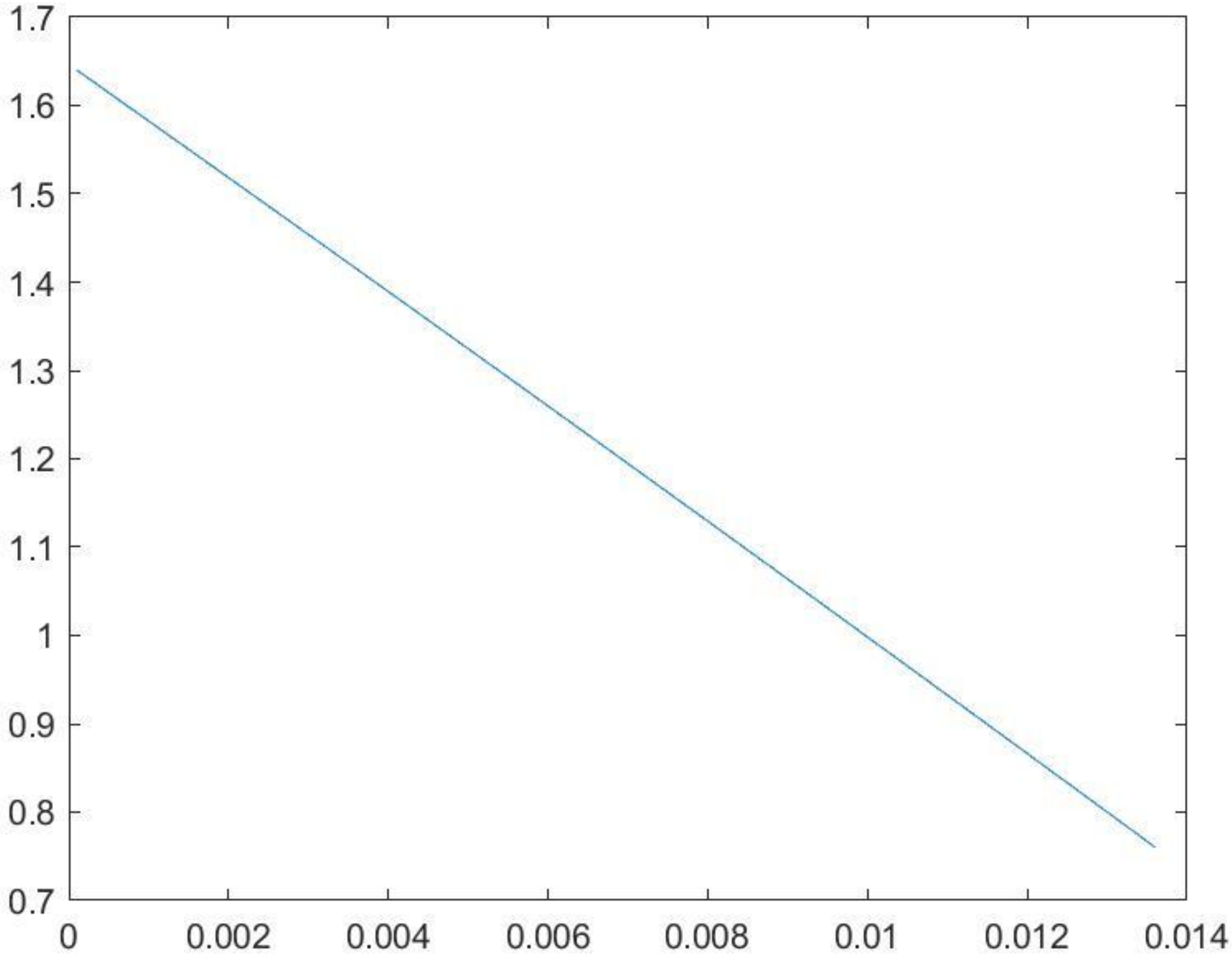

Figure 2. Dependence of the stratification-based Risk Ratio on p111 for the smoking confounder.

Among the admissible configurations, however, 26% have RR_ED_strat < 1.

Overall, the proposed ICA method gives the following result: Neither alcohol consumption (Sign_Conf = 16%) nor smoking (Sign_Conf = 26%) completely eliminates the positive association (E → D) in the majority of admissible configurations.

At the same time, smoking has a larger Significance of Confounder than alcohol consumption under the assumptions used here.

## 5.3 IVF Statistical Data

We next analyze in greater detail the IVF example introduced in Section 1, concerning the association between patient age and IVF success [2, 3]. Unlike the two previous examples, this dataset is particularly useful because the complete three-way statistical table is available. Consequently, the stratification-based Risk Ratio can be calculated directly and provides a reference value against which the reconstruction method can be evaluated.

The potential confounder is the oocyte source — whether the patient's own oocyte or a donor oocyte was used. We define:

- E — patient age, with E = 0 for ages 43 – 44 years and E = 1 for ages 45 – 50 years;
- U — oocyte source, with U = 0 for the patient's own oocyte and U = 1 for a donor oocyte;
- D — treatment outcome, with D = 0 for an unsuccessful outcome and D = 1 for a successful outcome.

Based on 8886 observations reported in [2], the complete three-way table is
N111 = 561, N110 = 1261, N101 = 49, N100 = 1385, N011 = 438, N010 = 779, N001 = 307, N000 = 4106.

The corresponding frequencies and observed Risk Ratios are pE = 0.366, pU = 0.342, pD = 0.152, RR_ED_obs =1.42, RR_EU_obs = 2.59, RR_UD_obs = 5.40.

Because all eight probabilities are known (Pijk = Nijk/8886), the stratification-based Risk Ratio can be calculated directly from equation (2):

RR_ED_strat = 0.76, compared with RR_ED_obs = 1.42. The apparent positive association E → D is therefore clearly explained by the oocyte-source confounder in this complete dataset.

Now suppose that the complete three-way table were unavailable and that only the frequencies and observed pairwise Risk Ratios were known.

The E-value calculation gives Bias = 2.00, and therefore RR_ED_strat >= RR_obs/Bias = 0.71.

The E-value analysis therefore permits the conclusion that the oocyte-source confounder **could explain** the observed association between patient age and IVF success. Again, however, this statement does not quantify how large a proportion of the admissible configurations actually leads to complete elimination of the association.

The proposed ICA method gives p111_low = 0.0284, p111_high = 0.0684, p111_root = 0.0300, Min_RR_ED_strat = 0.705, Max_RR_ED_strat = 1.009.

The dependence of the stratification-based Risk Ratio on p111 is shown in Fig. 3.

Because Max_RR_ED_strat > 1, complete elimination of the observed positive association E → D cannot be guaranteed for every admissible configuration. Nevertheless, independently of the assumed distribution of p111, RR_ED_strat <= Max_RR_ED_strat so the stratification-based effect cannot exceed 1.009. In particular, it is substantially smaller than the observed RR_ED_obs = 1.42.

Under the Uniform parameterization, the proposed method gives Sign_Conf = 97%. Thus, 97% of the admissible configurations correspond to complete elimination of the positive association E → D.

A Monte Carlo simulation with 100,000 runs, using independent uniformly distributed ±2% perturbations of the input parameters, gave MAE(Sign_Conf) = 3.9%. The main contribution to the deviation from the baseline value of 97% came from simulated values below 97%, since values above 97% are naturally bounded by 100%.

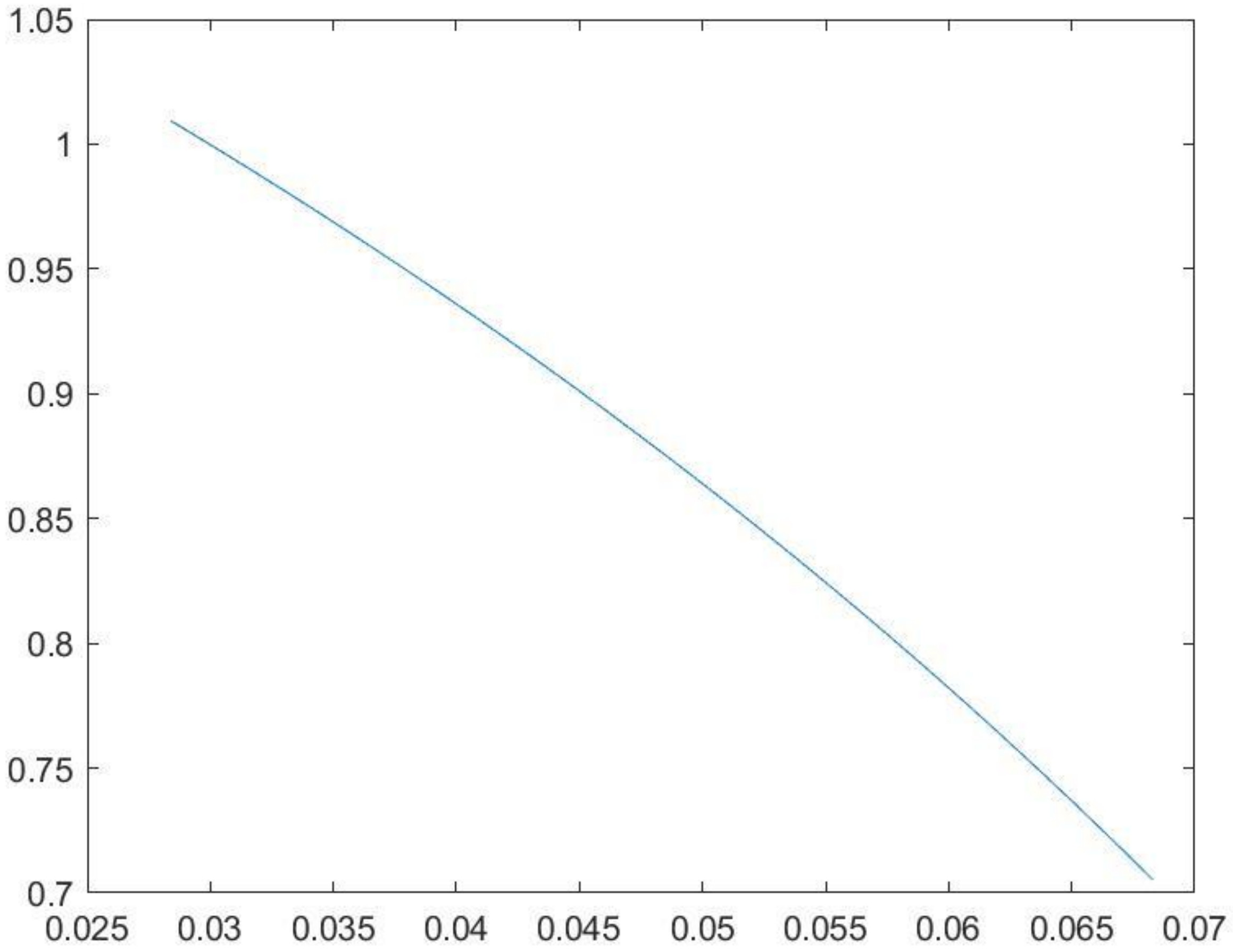

Figure 3. Dependence of the stratification-based Risk Ratio on p111 for the oocyte-source confounder.

## 6. Conclusion

A central problem in causal inference based on observational data is assessing the potential influence of unmeasured confounding on conclusions about the effect of an exposure on an outcome. The E-value is a widely used sensitivity-analysis tool for this purpose. However, its principal result is a one-sided bound on the possible confounding-induced distortion. In particular, when the corresponding Bias exceeds the observed Risk Ratio, the E-value analysis can only establish that a confounder **could** potentially explain the observed association. It does not quantify how frequently such an outcome occurs among the statistical configurations compatible with the observed characteristics.

The Inverse Confounding Analysis (ICA) method, proposed in this paper, addresses this broader question. The main result is an analytical description of the complete set of joint distributions compatible with specified frequencies and pairwise associations. The inverse reconstruction problem is formulated as a system of seven nonlinear equations in eight unknown probabilities. Although the original system is nonlinear, its complete solution set admits a simple linear parameterization by a single free parameter.

After imposing the probability constraints, the solution set becomes a one-dimensional admissible interval. Analytical expressions are obtained for both boundaries of this interval and for the corresponding stratification-based Risk Ratio. The latter is shown to be a fractional-linear and monotonic function of the free parameter. This analytical structure makes it possible to obtain two complementary types of information.

First, distribution-independent bounds can be calculated, including the minimum and maximum possible values of the stratification-based Risk Ratio. Second, when an assumed distribution of the free parameter p111 is specified as uniform, distribution-dependent measures can be calculated. In particular, we introduce Sign_Conf - proportion of admissible configurations for which RR_ED_strat < 1.

The proposed ICA method requires additional information compared with the E-value approach, namely the frequencies of the exposure, confounder, and outcome. Determining these frequencies may nevertheless be substantially easier than determining the corresponding pairwise Risk Ratios involving an unmeasured confounder.

The ICA method was illustrated using published observational data on antidepressant use during pregnancy and miscarriage [15] and using IVF data [2, 3]. The examples demonstrate that the statement that a confounder **could explain** an observed association may correspond to very different proportions of the admissible configuration space. In the examples considered here, the proposed method provides quantitative information about this distinction that is not supplied by the E-value bound alone.

The six-parameter specification used in the ICA method should not be regarded as the only possible formulation of the inverse reconstruction problem. Other sets of conditioning information may be considered depending on the data available in a particular application (see, for example, such data in [16] – Max and Min of conditional probabilities instead of Risk Ratio). Such extensions may lead to different admissible configuration spaces and, consequently, to different ranges of the stratification-based Risk Ratio. Investigation of alternative reconstruction schemes is an important direction for future work.

Natural extensions of the present Inverse Confounding Analysis method include:

- discrete confounders with more than two levels;
- continuous confounders;
- simultaneous influence of multiple confounders;